\documentclass[12pt,preprint]{aastex}

\def\um{$\mu$m }
\def\h2o{H$_2$O}
\def\ch4{CH$_4$}
\def\arcs{\ifmmode {''}\else $''$\fi}

\begin{document}
\title{IDENTIFYING YOUNG BROWN DWARFS USING GRAVITY-SENSITIVE SPECTRAL FEATURES}
\author{MARK R. MCGOVERN\altaffilmark{1},
J. DAVY KIRKPATRICK\altaffilmark{2},
IAN S. MCLEAN\altaffilmark{1},
ADAM J. BURGASSER\altaffilmark{1,3},
L. PRATO\altaffilmark{1},
PATRICK J. LOWRANCE\altaffilmark{2}}
\altaffiltext{1}{Department of Physics \& Astronomy, University of
California, Los Angeles, CA 90095-1562; mclean@astro.ucla.edu,
mcgovern@astro.ucla.edu, adam@astro.ucla.edu,
lprato@astro.ucla.edu}
\altaffiltext{2}{Infrared Processing and Analysis Center,
California Institute of Technology, Pasadena, CA 91125;
davy@ipac.caltech.edu, lowrance@ipac.caltech.edu}
\altaffiltext{3}{Hubble Fellow}

\begin{abstract}
We report the initial results of the Brown Dwarf
Spectroscopic Survey Gravity Project, to study gravity
sensitive features as indicators of youth in brown dwarfs.
Low-resolution (R$\sim$2000) $J$-band and optical (R$\sim$1000)
observations using NIRSPEC and LRIS at the W.M. Keck Observatory reveal
transitions of TiO, VO, K~I, Na~I, Cs~I, Rb~I, CaH, and FeH. By comparing
these features in late-type giants and in old field dwarfs we show that they
are sensitive to the gravity ($g=GM/R^{2}$) of the object. Using
low-gravity spectral signatures as age indicators, we observed
and analyzed $J$-band and optical spectra of two young brown dwarfs,
G 196-3B (20-300 Myr) and KPNO Tau-4 (1-2 Myr), and two possible low
mass brown dwarfs in the $\sigma$ Orionis cluster (3-7 Myr).
We report the identification of the $\phi$ bands of TiO near 1.24 \um
and the A-X band of VO near 1.18 \um together with extremely weak
$J$-band lines of K I in KPNO-Tau4.  This is the first detection of TiO
and VO in the $J$-band in a sub-stellar mass object.
The optical spectrum of KPNO-Tau4 exhibits weak K~I and Na~I lines, weak absorption
by CaH, and strong VO bands, also signatures of a lower gravity atmosphere.
G 196-3B shows absorption features in both wavelength regions like those of
KPNO-Tau4 suggesting that its age and mass are at the lower
end of published estimates. Whereas $\sigma$ Ori 51 appears to be consistent
with a young sub-stellar object, $\sigma$ Ori 47 shows
signatures of high gravity most closely resembling an old L1.5/L0,
and can not be a member of the $\sigma$ Orionis cluster.
\end{abstract}


\section{INTRODUCTION}
Theoretical models predict that sub-stellar objects are much more
luminous and hotter when very young (Burrows et al.\ 2001).
Consequently, deep imaging surveys of young clusters can yield
very low mass sub-stellar objects that will be too faint to study
once they have aged, unless they are located extremely close to
the Sun. Hence, the very low end of the initial mass function can be
more effectively probed in young clusters. Several groups have reported deep
infrared surveys of young star-forming regions at distances
ranging from 140-450 pc (Luhman et al.\ 1998, 1999ab, 2000ab;
B\'{e}jar et al.\ 1999; Stauffer et al.\ 1999; Ardila et al.\ 2000;
Lucas \& Roche 2000; Zapatero Osorio et al.\ 2000; Mainzer \&
McLean 2003, among others). For example, in the Sigma Orionis
open cluster (4.2$^{+2.7}_{-1.5}$ Myr, Oliveira et al.\ 2002;
352$^{+166}_{-85}$ pc, Hipparcos), Zapatero Osorio et al.\ (2000)
report several sub-stellar mass objects which, if members of the
association, would have masses in the range 8-15 M$_{jupiter}$.

One way to distinguish between young and old brown dwarfs is to
look for gravity-sensitive spectral features. The radius of field
brown dwarfs varies only slightly with mass and age, and therefore
the surface gravity ($g=GM/R^{2}$) is determined by the mass (log
g$\sim$5). At a young evolutionary stage however, lower mass
objects will have effective temperatures corresponding to those of
late M or early L dwarfs. Furthermore, the radius of these young
objects can be as much as three times greater than their eventual
equilibrium state (Burrows et al.\ 2001). As a result, young
objects can exhibit significantly lower surface gravities (10-100
times) than the more massive evolved dwarfs of the same spectral
type. Mart{\'{\i}}n et al.\ (1996) and Luhman et al.\ (1997)
demonstrated that CaH, K I, Na I, and VO features can be used as
gravity-sensitive diagnostics in the optical spectra of brown
dwarf candidates. Gorlova et al.\ (2003) showed that the neutral
potassium (K I) lines in the $J$-band are also sensitive to
surface gravity.

In this current paper we present the first results from our survey that
focus on young brown dwarfs and their surface gravity spectral
indicators. We show that for ages $\lesssim$5 Myr, not only are
the K I lines and the FeH feature in the $J$-band very weak for a
given spectral type, but also infrared TiO and VO bands, normally not
present at the boundary between M and L dwarfs in the old field population,
are present for the lowest mass objects. We also reveal that the optical
spectra of very young brown dwarfs exhibit weaker lines of K I, Na I,
Rb I, and Cs I; weaker bands of CaH; and stronger bands of VO than
field dwarfs of the same spectral class. These results, and the
extensive spectral data base of old field M, L (Kirkpatrick et al.\ 1999),
and T (Burgasser et al.\ 2002) dwarfs in the Brown Dwarf Spectroscopic Survey
(BDSS, McLean et al.\ 2003; hereafter M03) and the Kirkpatrick
L Dwarf Archive \footnote{\url{http://spider.ipac.caltech.edu/staff/davy/ARCHIVE/}}
enable us to test the membership of potential
very low mass brown dwarfs in young clusters. We have applied these
criteria to KPNO-Tau4 in the Taurus-Auriga region (Kenyon et al.\ 1994a,
Brice\~{n}o et al.\ 2002), G 196-3B (Rebolo et al.\ 1998),
and $\sigma$ Ori 47 and $\sigma$ Ori 51 in the Sigma Orionis open cluster
(B\'{e}jar et al.\ 1999, Zapatero Osorio et al.\ 2000). In \S 2 we describe
the observations and methods. Results are presented and discussed in \S3, and
our conclusions are summarized in \S4.

\section{OBSERVATIONS AND RESULTS}
The objects observed and their relevant photometric properties are listed
in Table 1. KPNO-Tau4 was discovered by Brice\~{n}o et al.\ (2002) as
part of an $I$ and $z$$^{\prime}$ deep survey of 8 square degrees in the
Taurus-Auriga dark cloud complex (1-2 Myr; 140$\pm$10 pc, Kenyon
et al.\ 1994b). An optical spectrum obtained by the authors was used
to assign a spectral type of M9.5$\pm$0.5. G 196-3B was discovered by
Rebolo et al.\ (1998) from a direct imaging search around young, nearby,
cool dwarf stars. Its age was estimated at 20-300 Myr based upon the
chromospheric and coronal properties of G 196-3A. Optical spectra indicate
G 196-3B is an L2 dwarf with a mass estimated at 25$^{+15}_{-10}$ Jupiter
masses (Rebolo et al.\ 1998). $\sigma$ Ori 47 was first reported as a
sub-stellar candidate member of the $\sigma$ Orionis cluster (B\'{e}jar
et al.\ 1999) as part of a deep $RIz$ survey in an 870 square arcminute
field around the O9.5V star $\sigma$ Orionis. Follow up infrared photometry
and optical spectroscopy acquired by Zapatero Osorio et al.\ (1999)
suggested a substellar nature of $\sigma$ Ori 47.
Zapatero Osorio et al.\ presented a Keck LRIS spectrum of this object and
concluded that the strengths of the hydride and alkali lines are lower than
in a field L dwarf of the same type. This would indicate lower gravity and
hence a young age for $\sigma$ Ori 47. They also claimed a lithium detection
of 4.3$\pm$0.5 \AA~EW, another indication of youth. A spectral type of
L1.5$\pm$0.5 was assigned based upon the optical spectrum. $\sigma$ Ori
51 (Zapatero Osorio et al.\ 2000) was discovered during a deeper follow
up survey of the initial B\'{e}jar et al.\ (1999) sample. Sub-stellar status
was determined from fits of the color data to model isochrones for young
brown dwarfs. Subsequent low resolution optical spectroscopy was conducted
by Barrado y Navascu\'{e}s et al.\ (2001) who determined its spectral type
to be M9.0$\pm$0.5. Finally, and most importantly, we observed the Mira
variable IO Virginis (M9.5 III+) as a template for very low
gravities (log g$\sim$0).

Near-infrared (near-IR) observations were made with NIRSPEC
on the Keck II telescope, using a 0.43\arcsec~ slit to yield a
resolution of R$=\lambda/\Delta\lambda \sim$ 2000 at $J$-band
(1.14-1.36 \micron). The $J$-band was chosen because it contains
four strong lines of neutral potassium, FeH bands, and a strong
\h2o band useful for spectral classification (M03). A detailed
description of NIRSPEC is given elsewhere (McLean et al.\ 1998,
2000). Observations were made as a sequence of nodded pairs by
positioning the object at two locations along the slit separated
by $\sim$20\arcsec. Each exposure was 300 s and at least two
nodded pairs were obtained for each source, giving total
integration times of 20-60 minutes. The typical signal-to-noise ratio
per resolution element for these observations is 20-30, with the
exception of $\sigma$ Ori 51 which has a signal-to-noise of
$\sim$10. Data extraction and reduction techniques were carried
out using REDSPEC, the NIRSPEC reduction package (see M03
for details).

Optical spectra were obtained at the Keck I telescope using the Low
Resolution Imaging Spectrograph (LRIS; Oke et al. 1995). For most of the
observations listed in Table 1, the setup and reduction procedures were
identical to those discussed in Kirkpatrick et al. (1999).
The only exception is the spectrum of G 196-3B, which is pieced together from
two different LRIS setups. The first setup was used on 1999 Mar
04/05, resulting in an 9{\AA}-resolution spectrum of G 196-3B from 6300 to
10000 \AA. On 2001 Feb 19 we used the 300 lines/mm grating
blazed at 5000 \AA~to cover the interval from 3800 to 8500 \AA. Reductions
were identical, and the resulting spectrum has a slightly lower resolution of
11 \AA. Because the blue spectrum was taken in better sky conditions, we
have constructed a 6300-10000 \AA~spectrum for G 196-3B by combining both
data sets: 1999 Mar data for $\lambda$ $>$ 8500 \AA~and 2001 Feb data for
$\lambda$ $<$ 8500 \AA. Spectra obtained on 2003 Jan 02/03 and 2001 Feb 19
have been corrected for telluric absorption. This was accomplished by
observing a G0 dwarf at similar airmass shortly before or after the target
observation and interpolating the continuum across the telluric bands to
determine the correction.

\section{DISCUSSION}
Figure 1 shows the NIRSPEC $J$-band spectrum of IO Vir compared
with that of KPNO-Tau4 and G 196-3B. A remarkable feature in the $J$-band
spectra is the appearance of a broad absorption band near 1.24 \um
in both the Mira variable IO Vir and the
young M9.5 dwarf KPNO-Tau4. These bands are probably not the result
of FeH, as related features at 1.20 \um are very weak or absent.
We instead identify these features as the $\phi$ ($\Delta\nu=-1$)
bandheads of TiO (Galehouse et al.\ 1980). A second feature near
1.18 \um is attributed to the A-X ($\Delta\nu=-1$) band of VO
(Cheung et al.\ 1982). These oxide bands are common in the late type $J$-band
spectra of Mira variables (Hinkle et al.\ 1989; Joyce et al.\ 1998) and
apparently also present in the low gravity atmospheres of young brown
dwarfs. The K~I doublets (1.168,1.177 \um and 1.243,1.254 \micron) are
absent in the cool, low-gravity atmosphere of IO Vir and only weakly
present in the spectrum of KPNO-Tau4, consistent with the behavior noted by
Gorlova et al.\ (2003). The \h2o band at 1.34 \um in KPNO-Tau4 is deeper
than expected for an M9.5 object, being more consistent with a
L2 dwarf (M03). This effect could likely be the result of the \h2o band
showing some slight sensitivity to gravity, as is the case in giant star
atmospheres, but the effect is much less noticable than for the K~I lines.
The spectrum of the L2 dwarf G 196-3B exhibits
similar features. The deep absorption in the range 1.14-1.20 \um can be
attributed to a mixture of VO, TiO and possibly \h2o. We have
confirmed the identification of the TiO and VO bands by observing these
objects at shorter wavelengths ($z$-band) where the fundamental transitions
($\Delta\nu=0$) are easily detected. We conclude that the presence of TiO \&
VO, weak K I lines and weak FeH absorption in the near-IR spectra of KPNO-Tau4 and
G 196-3B are indicative of low surface gravity and hence support their
identification as young, low-mass brown dwarfs.

Figure 2 shows LRIS spectra of IO Vir, KPNO-Tau 4, 2MASSW
J0149090+295613 (hereafter, 2M0149+29; from Liebert et al. 1999),
G 196-3B, and Kelu-1 (from Kirkpatrick et al. 1999). The top three
spectra show the effects of differing gravity in late M stars. The
M giant IO Vir exhibits the hallmarks of low-gravity, with very
weak or non-existent absorptions by K I and Na I, weak absorption
by CaH, and strong, bowl-shaped depressions by VO. The presumably
solar-age field dwarf 2M0149+29 (M9.5 V) shows the hallmarks of
higher gravity -- strong absorptions by K I and Na I, strong
absorption by CaH, and less pronounced features attributable to
VO. The spectrum of KPNO-Tau 4 shows spectroscopic characteristics
intermediate between the field dwarf and the giant (cf.
Brice{\~n}o et al. 2002). Such a gravity signature is expected for
a dwarf that is much less massive and/or more extended than a
higher mass field counterpart of similar temperature. This result
is consistent with KPNO-Tau 4 being young.

The bottom two spectra of Figure 2 show the young L2 dwarf G 196-3B compared
to the presumably older field L2 dwarf Kelu-1. Again the spectrum of the
younger (lower-mass) object shows the signatures of lower gravity, most
notably the weaker lines of the alkali elements K I, Rb I, Na I, and Cs I.
The Li I line shows the opposite effect, as lithium has been largely
depleted in the more massive Kelu-1, but is retained in the lower mass brown
dwarf G 196-3B (Rebolo et al. 1992).

We now consider the $\sigma$ Orionis objects. Based on the near-IR and
optical signatures of low gravity established in Figures 1 and 2, the
comparison of the $J$-band and optical spectra of $\sigma$ Ori 47 and 51 in
Figures 3 and 4, respectively, is revealing. The $J$-band spectrum of $\sigma$ Ori 47
contains strong K~I lines, moderate FeH absorption at 1.20 \um
and relatively deep \h2o absorption at 1.34 \micron. On the other hand,
the $J$-band spectrum of $\sigma$ Ori 51 has very weak K~I lines and very
weak FeH absorption. Although the TiO and VO bands are not detected, the
weakness of the K~I lines and FeH band in the $J$-band spectrum of $\sigma$
Ori 51 is consistent with it being a lower gravity object. A comparison
of the \h2o absorption band at 1.34 \um to the BDSS sample (M03) shows that the
depth of this band matches most closely with an L0-L1 dwarf for $\sigma$ Ori
51. Again, a slightly later IR spectral type suggests that, at our
resolution of R$\sim$2000, we could likely be detecting some sensitivity
of the \h2o band to gravity.  The absence of the $J$-band TiO and VO bands
in $\sigma$ Ori 51 may be attributed to its somewhat higher surface gravity
as compared to KPNO-Tau4 because of its older age ($\sim$5 Myr compared
to $\sim$1 Myr) and warmer atmosphere although veiling is a viable
alternative hypothesis. The strength of the near-IR spectral features
observed in $\sigma$ Ori 47 are consistent with the known field dwarf
2MASS 0345+35 (L0; Kirkpatrick et al.\ 1997a) plotted at the top of
Figure 3 for comparison. Based on the equivalent width of
the K I lines and the strength of the \h2o band at 1.34 \micron, the
near-IR data support the conclusion that $\sigma$ Ori 47 appears
to be a much older L0$\pm$0.5 dwarf.

Figure 4 shows LRIS spectra for these objects. A comparison of the
spectrum of $\sigma$ Ori 51 to the M star spectra in Figure 2
confirms that it has the signatures of lower gravity, most notably
weak Na I, bowl-shaped depressions by VO, and weak CaH,
demonstrating that it is a young object. For $\sigma$ Ori 47, we
find that the opposite is true, in contradiction to the claims of
Zapatero Osorio et al.\ (1999). Inset (A) in Figure 4 shows a
comparison of the spectra of G 196-3B (from Figure 2) and $\sigma$
Ori 47 in the region around the 6708 \AA~Li I line. We find no
lithium line down to 2 \AA~EW (Li I in G 196-3B has EW=6 \AA).
Inset (B) shows a comparison of $\sigma$ Ori 47 to the low-gravity
dwarf G 196-3B and to the presumably high-gravity L1.5 V field
dwarf 2MASSW J0832045$-$012835 (hereafter, 2M0832-01; Kirkpatrick
et al. 2000) in the region near the other alkali lines. The
$\sigma$ Ori 47 spectrum shows a strong Na I doublet at 8183 and
8195 \AA~like that of a field L dwarf. Fortunately, this Na I
doublet falls in a region with very little contamination by
telluric OH emission, so the spectrum here is less noisy. Even in
the wavelength regions where residual OH emission causes more of a
problem, such as near the Rb I and Cs I lines at 7800, 7948, and
8521 \AA, absorptions by the alkali elements are obvious and of
comparable strength to the absorptions in a field L dwarf.

We classify $\sigma$ Ori 47 in the optical as an L1.5, the
same spectral type assigned to 2M0832-01, and suggest that within the
uncertainties the two spectra are identical. Hence $\sigma$ Ori 47 appears to
be an old, high-mass, field L dwarf. The L1.5 optical spectral type and 2MASS J
magnitude of 17.53$\pm$0.24 suggest a distance of 120 pc (see eq. 1 from
Kirkpatrick et al. 2000), roughly a factor of three closer than the cluster
itself. The higher gravity
implies an age closer to that of the Sun, which means the mass of this L dwarf
is near the stellar/substellar break of 0.075-0.080 $M_{\odot}$ (Burrows et al.
1997). We conclude that this object shows no signs of youth, is unassociated
with the cluster, and is not an object ``reaching the mass boundary
between brown dwarfs and giant planets'' near 0.015 $M_{\odot}$ as originally
reported (Zapatero Osorio et al. 1999).

\section{CONCLUSIONS}
The identification of gravity-sensitive spectral features is a
powerful technique in the study of brown dwarfs. By comparison
with a known object of lower surface gravity (log g$\sim$0),
several indicators of low gravity in brown dwarfs have been
identified. We have shown empirically that in the $J$-band the
$\phi$ ($\Delta\nu=-1$) bandheads of TiO and the A-X
($\Delta\nu=-1$) band of VO appear in very young late-type M or
early L objects as a result of low surface gravity. This is the
first detection of these bands in a sub-stellar object. In
slightly older brown dwarfs weak FeH and K~I features indicate low
gravity. Absorption from \h2o in the $J$-band is slightly greater
than expected for the given spectral type. In optical spectra,
weak alkali lines of Na~I, K~I, Rb~I and Cs~I, but strong Li~I
absorption, together with weak CaH and strong VO absorption are
indicative of low gravity. Older late-M and early-L dwarfs with
higher surface gravity always exhibit strong alkali lines in both
the visible and near-IR, strong FeH in the $J$-band, and strong
CaH around 7000 \AA. For KPNO-Tau4 and $\sigma$ Ori 51 we find
optical and infrared spectral features consistent with low gravity
and youth, a necessary condition for membership in a young
cluster. G 196-3B is also confirmed as a young brown dwarf,
consistent with it being coeval with its more massive companion.
$\sigma$ Ori 47 however, does not exhibit any of the features of
low surface gravity and we conclude that it is not young and
therefore not a low-mass member of the $\sigma$ Orionis cluster,
but instead an early-L field dwarf with a mass near the
stellar/substellar boundary.

\acknowledgments We wish to thank the staff of the Keck Observatory for
their support. This research has made use of the NASA/IPAC Infrared Science Archive,
which is operated by the Jet Propulsion Laboratory, California Institute of Technology,
under contract with the National Aeronautics and Space Administration.
This publication makes use of data products from the Two Micron All Sky Survey, which
is a joint project of the University of Massachusetts and the Infrared Processing
and Analysis Center, funded by the National Aeronautics and Space Administration and the
National Science Foundation. A. J. B. acknowledges support by NASA through Hubble
Fellowship grant HST-HF-01137.01, awarded by the Space Telescope Science Institute,
which is operated by the Association of Universities for Research in Astronomy, Inc.,
for NASA, under contract NAS 5-26555. The authors are also grateful to those of
Hawaiian ancestry on whose sacred mountain we are privileged to be guests.

\clearpage

\begin{figure}[!htp]
\epsscale{0.95}
\plotone{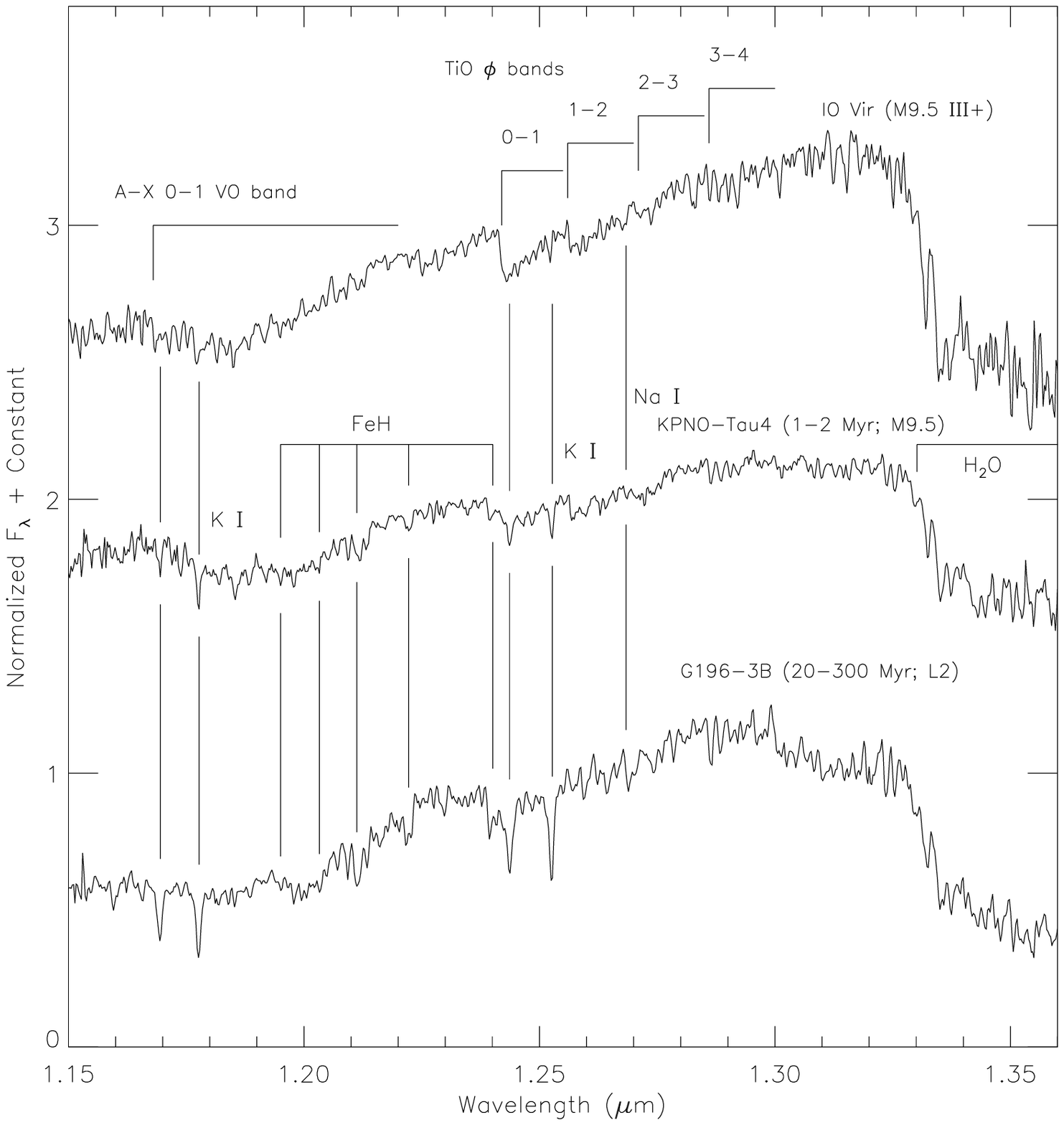} \caption{The NIRSPEC $J$-band
spectra of the young, low-mass brown dwarfs KPNO-Tau4 and G 196-3B
are plotted together with the late-M giant IO Vir. The spectrum of
IO Vir shows absorption bands of VO and TiO located at 1.18 \um and
1.24-1.28 \micron, respectively. Similar features appear in the
spectrum of KPNO-Tau4 and G 196-3B suggesting that they have low gravity
atmospheres. Spectral types listed here are based upon optical
classifications. Spectra are normalized at 1.265 \um and separated from
one another vertically by integer offsets.}
\end{figure}
\epsscale{1}

\begin{figure}[!htp]
\rotatebox{-90}{
\epsscale{0.8}
\plotone{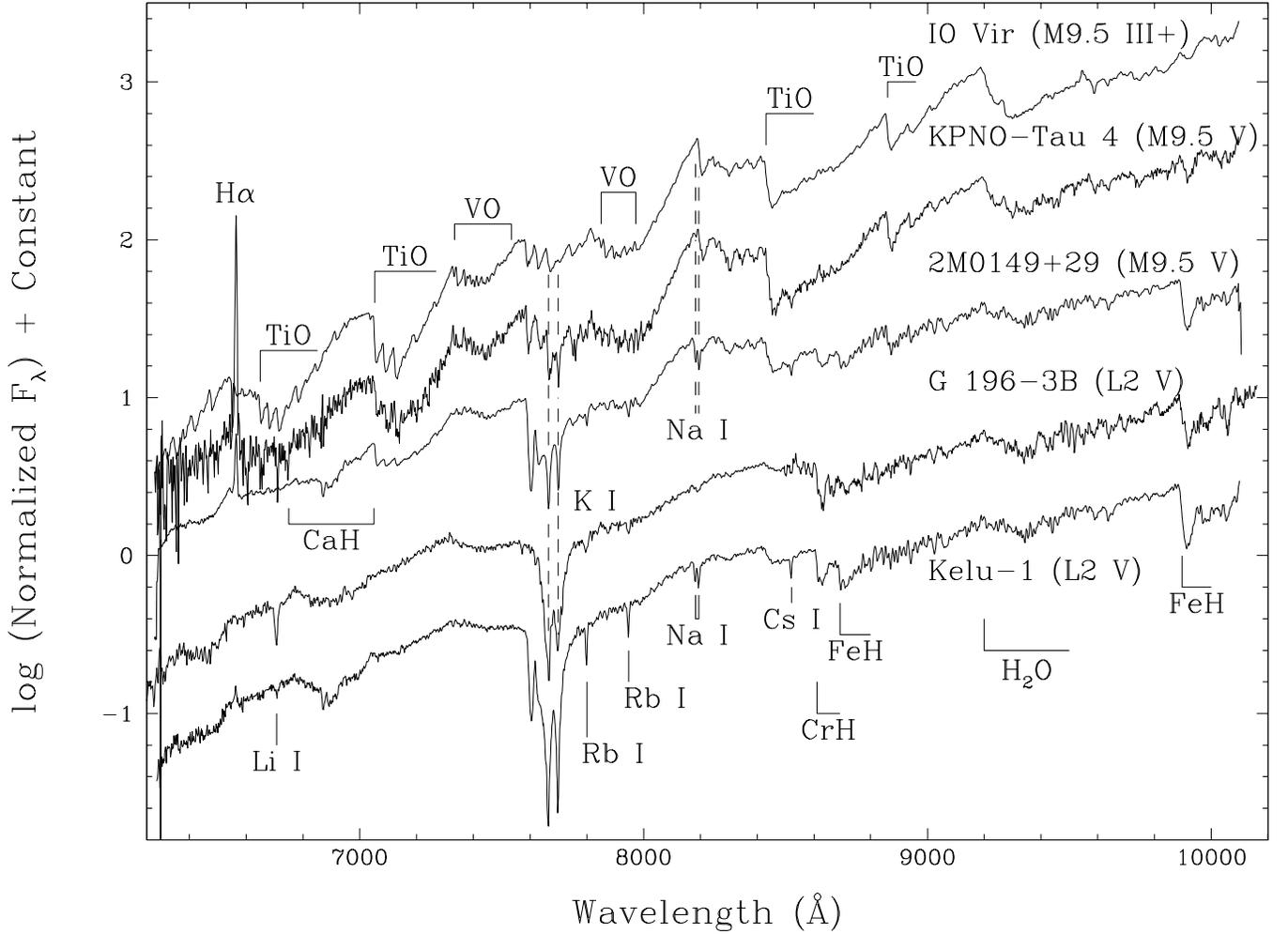}}
\caption{LRIS optical spectra of three M9.5 objects (top) and
two early L dwarfs (bottom). Noteworthy spectral features are
labelled. As noted in the text, the telluric A-band (6867-7000
\AA) and B-band (7594-7685 \AA) of $O_2$ are evident in the
spectra of 2M0149+29 and Kelu-1 but have been corrected in the
other spectra. Note the differences in feature strengths in
the M9.5 spectra from the low-gravity giant IO Vir down to
the high-gravity field dwarf 2M0149+29, and between the lower
gravity, young L dwarf G 196-3B and the higher gravity, field
L dwarf Kelu-1. See text for more details.}
\end{figure}
\epsscale{1}

\begin{figure}[!htp]
\epsscale{0.95}
\plotone{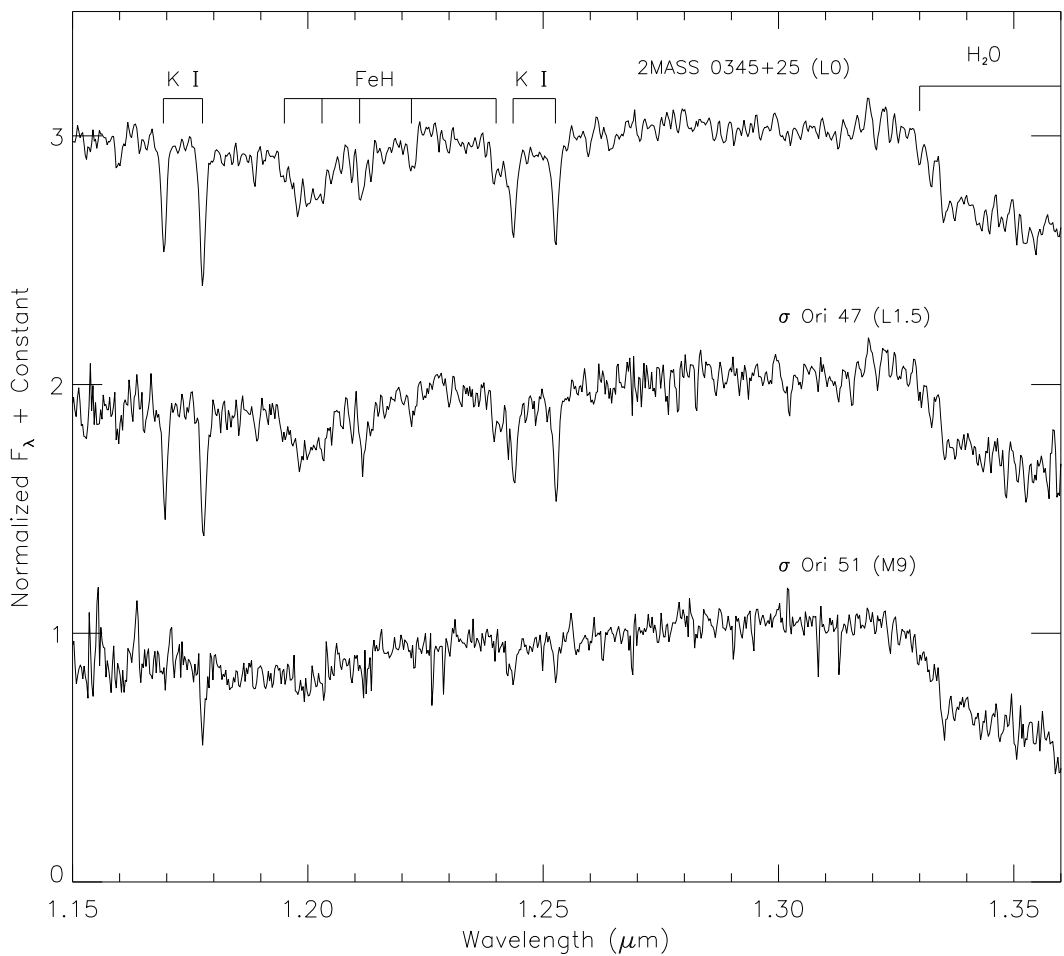} \caption{NIRSPEC $J$-band spectra
of 2MASS 0345+25, $\sigma$ Ori 47, and $\sigma$ Ori 51. 2MASS 0345+25
is plotted for comparison as a known higher gravity object. $\sigma$
Ori 51 exhibits very weak K I lines and weak FeH absorption, consistent
with a low surface gravity, whereas $\sigma$ Ori 47 has relatively
strong absorption features at these locations and, at these wavelengths,
matches a field L0 dwarf most closely. Spectral types listed here are based
upon optical classifications. Spectra are normalized at 1.265 \um and
separated from one another vertically by integer offsets.}
\end{figure}
\epsscale{1}

\begin{figure}[!htp]
\rotatebox{-90}{ \epsscale{0.75} \plotone{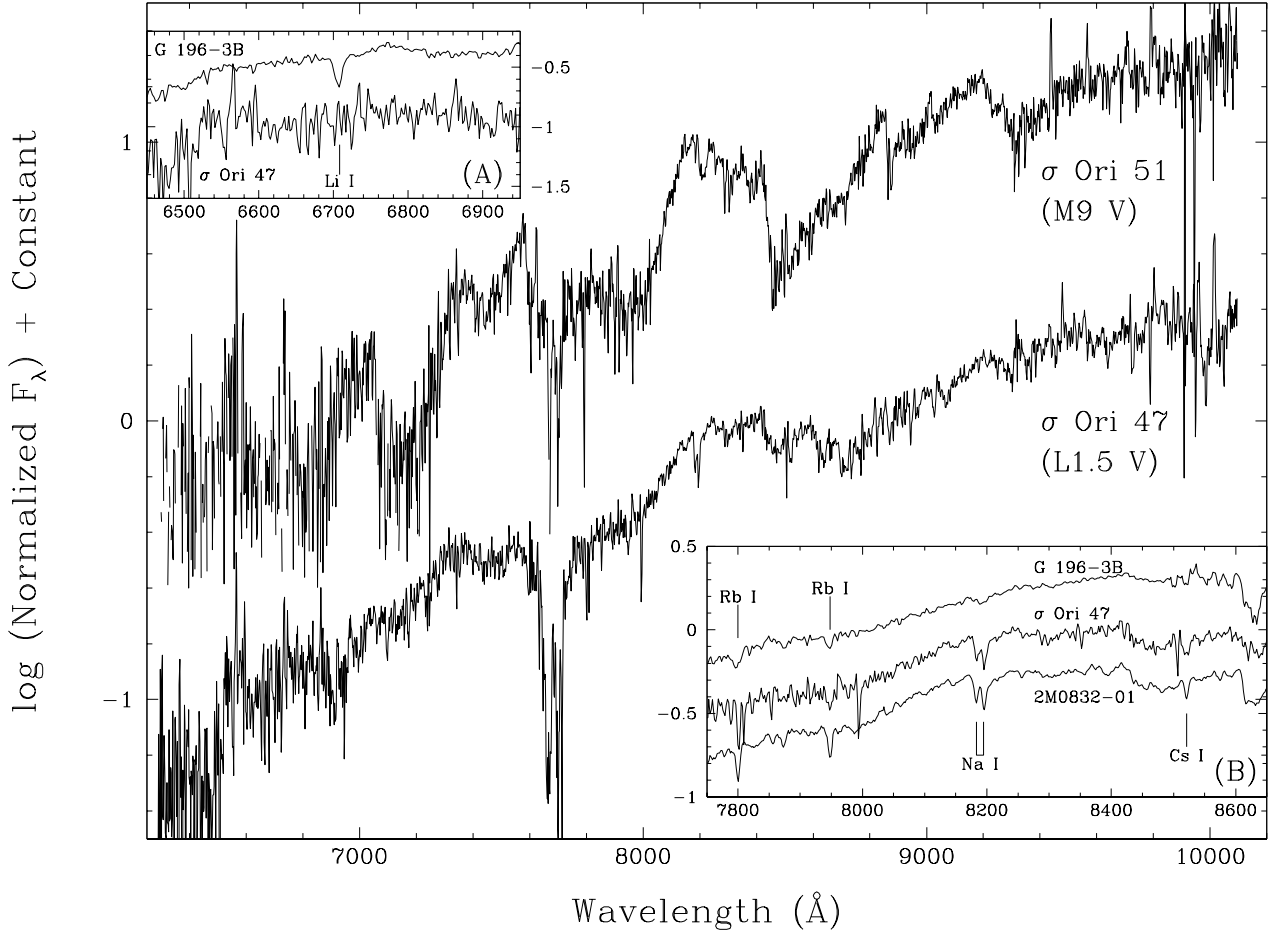}} \caption{LRIS
optical spectra of two candidate members of the $\sigma$ Orionis
cluster. Both spectra have been corrected for telluric absorption,
and feature identifications can be found in Figure 2. Comparison
to the spectra in Figure 2 further shows that $\sigma$ Ori 51 is a
low-gravity object and $\sigma$ Ori 47 is not. This latter point
is further demonstrated in insets A and B. Inset A illustrates no
obvious Li I absorption line, as the comparison to the low-gravity
spectrum of G 196-3B (L2 V) shows. Inset B shows that the alkali
line strengths of $\sigma$ Ori 47 are similar to the field dwarf
2M0832-01 (L1.5 V) but quite unlike the alkali strengths in the
young G 196-3B. See text for more details.}
\end{figure}
\epsscale{1}

\clearpage

\begin{deluxetable}{lcccccccccc}
\tabletypesize{\scriptsize} \tablecaption{Observing Log and
Properties of Sources} \tablewidth{0pt}
\tablehead{\colhead{Object} & \colhead{Spectral} & \colhead{R.A.} & \colhead{Decl.} &
\colhead{} & \colhead{} & \colhead{UT Date} & \colhead{UT Date} & \colhead{} \\ \colhead{} & \colhead{Type} & \colhead{(J2000.0)} & \colhead{(J2000.0)} & \colhead{$I$(mag)} & \colhead{$J$(mag)} & \colhead{Optical Obs.} & \colhead{NIR Obs.} & \colhead{Refs}}
\startdata
$\sigma$ Ori 47 & L1.5/L0 V & 05 38 14.5 & -02 40 16 & 20.5 & 17.53 & 2003 Jan 03 & 2003 Mar 24    & 1,2,9 \\
$\sigma$ Ori 51 & M9 V    & 05 39 03.2 & -02 30 20 & 20.7 & 17.21 & 2003 Jan 02 & 2002 Dec 24    & 3     \\
KPNO-Tau4       & M9.5 V    & 04 27 28.0 & +26 12 05 & 18.7 & 15.00 & 2003 Jan 03 & 2002 Dec 24    & 4,9   \\
G 196-3B        & L2 V      & 10 04 20.7 & +50 23 00 & 18.3 & 14.90 & 1999 Mar 04/05 & 2001 Mar 06 & 5,6,9 \\
                &           &            &           &      &       & 2001 Feb 19 &                &       \\
IO Vir          & M9.5 III+ & 14 11 17.5 & -07 44 50 & 12.1 & 6.65  & 2003 Jan 03 & 2003 Mar 24    & 7,8,9 \\
\enddata
\tablerefs{(1) B\'{e}jar et al.\ 1999; (2) Zapatero Osorio et al.\ 1999; (3) Zapatero Osorio et al.\ 2000; (4) Brice\~{n}o et al.\ 2002; (5) Rebolo et al.\ 1998; (6) Kirkpatrick et al.\ 2001; (7) Kirkpatrick et al.\ 1997b; (8) USNO-B1.0 Catalog; (9) 2MASS catalog.}
\end{deluxetable}


\begin{references}
Ardila, D., Mart{\'{\i}}n, E., \& Gibor, B. 2000, AJ, 120, 479 \\
Barrado y Navascu\'{e}s, D., et al. 2001, A\&A, 377, L9 \\
B\'{e}jar, V.J.S., Zapatero Osorio, M.R., \& Rebolo, R. 1999, ApJ, 521, 671 \\
Brice\~{n}o, C., Luhman, K.L., Hartmann, L., Stauffer, J.R., \& Kirkpatrick, J.D. 2002, ApJ, 580, 317 \\
Burgasser, A.J., et al. 2002, ApJ, 564, 421 \\
Burrows, A., et al. 1997, ApJ, 491, 856 \\
Burrows, A., Hubbard, W.B., Lunine, J.I., \& Liebert, J. 2001, Rev. Mod. Phys., 73, 719 \\
Cheung, A.S.-C., Taylor, A.W., \& Merer, A.J. 1982, J. Mol. Spectrosc., 68, 399 \\
Dahn, C.C., et al. 2002, AJ, 124, 1170 \\
Galehouse, D.C., Brault, J.W., \& Davis, S.P. 1980, ApJ, 42, 241 \\
Gorlova, N., Meyer, M.R., Liebert, J. \& Rieke, G.H. 2003, astro-ph/0305147  \\
Hinkle, K.H., Lambert, D.L., \& Wing, R.F. 1989, MNRAS, 238, 1365 \\
Joyce, R.R., Hinkle, K.H., Wallace, L., Dulick, M., \& Lambert, D.L. 1998, AJ, 116, 2520 \\
Kenyon, S.J., Gomaz, M., Marzke, R.O., \& Hartmann, L. 1994a, AJ, 108, 251 \\
Kenyon, S.J., Dobrzycka, D., \& Hartmann, L. 1994b, AJ, 108, 1872 \\
Kirkpatrick, J. D., Beichman, C. A., \& Skrutskie, M. F. 1997a, ApJ, 476, 311 \\
Kirkpatrick, J. D., Henry, T.J., \& Irwin, M.J. 1997b, AJ, 113, 1421 \\
Kirkpatrick, J. D., et al. 1999, ApJ, 519, 802 \\
Kirkpatrick, J. D., et al. 2000, AJ, 120, 447 \\
Kirkpatrick, J. D., et al. 2001, AJ, 121, 3235 \\
Liebert, J., Kirkpatrick, J. D., Reid, I. N., \& Fisher, M. D. 1999, ApJ, 519, 345 \\
Lucas, P.W. \& Roche, P.F. 2000, MNRAS, 314, 858 \\
Luhman, K.L., Liebert, J., \& Rieke, G. H. 1997, ApJL, 489, 165 \\
Luhman, K.L., Rieke, G.H., Lada, C.J., \& Lada, E.A. 1998, ApJ, 508, 347 \\
Luhman, K.L., \& Rieke, G.H. 1999a, ApJ, 525, 440 \\
Luhman, K.L., 1999b, ApJ, 525, 466 \\
Luhman, K.L., et al. 2000a, ApJ, 543, 299 \\
Luhman, K.L., 2000b, ApJ, 544, 1044 \\
Mainzer, A.K. \& McLean, I.S. 2003, ApJ, in press \\
Mart{\'{\i}}n, E. L., Rebolo, R., \& Zapatero Osorio, M. R. 1996, ApJ, 469, 706 \\
McLean, I.S., et al. 1998, Proc. SPIE, 3354, 566 \\
McLean, I.S., et al. 2000, Proc. SPIE, 4008, 1048 \\
McLean, I.S., McGovern, M.R., Burgasser, A.J., Kirkpatrick, J.D., Prato, L., \& Kim, S.S. 2003, ApJ, 596, in press (M03) \\
Oke, J.B., et al. 1995, PASP, 107, 375 \\
Oliveira, J.M., Jeffries, R.D., Kenyon, M.J., Thompson, S.A., \& Naylor, T. 2002, A\&A, 382, L22 \\
Rebolo, R., Mart{\'\i}n, E. L., \& Magazz{\'u}, A. 1992, ApJL, 389, 83 \\
Rebolo, R., Zapatero Osorio, M.R., Madruga, S., B\'{e}jar, V.J.S., Arribas, S., \& Licandro, J. 1998, Science, 282, 1309 \\
Stauffer, J.R., et al. 1999, ApJ, 527, 219 \\
Zapatero Osorio, M.R., B\'{e}jar, V.J.S., Rebolo, R., Mart{\'{\i}}n, E.L., \& Basri, G. 1999, ApJL, 524, 115 \\
Zapatero Osorio, M.R., B\'{e}jar, V.J.S., Mart{\'{\i}}n, E.L., Rebolo, R., Barrado Y Navascu\'{e}s, D., Bailer-Jones, C.A.L., \& Mundt, R. 2000, Science, 290, 103 \\
\end{references}
\end{document}